\documentclass[preprint,authoryear,12pt]{elsarticle}
\usepackage{amsthm}
\usepackage[show]{ed}

\journal{Progress in Biophysics \& Molecular Biology}

\title{A Quantum Model of Exaptation: Incorporating Potentiality into Evolutionary Theory}
\author[ubc]{Liane Gabora}
\ead{liane.gabora@ubc.ca}
\author[gmu]{Eric O. Scott}
\author[uvm]{Stuart Kauffman}
\address[ubc]{Department of Psychology, University of British Columbia, Okanagan Campus, 3333 University Way, Kelowna, British Columbia, V1V 1V7, Canada}
\address[gmu]{Department of Computer Science, George Mason University, 400 University Drive MSN 4A5, Fairfax, VA 22030}
\address[uvm]{Department of Mathematics and Statistics, University of Vermont, 16 Colchester Ave., Burlington, VT 05401}

\date{}

\begin{document}
\begin{frontmatter}
\begin{abstract}
The phenomenon of preadaptation, or \emph{exaptation} (wherein a trait that originally evolved to solve one problem is co-opted to solve a new problem) presents a formidable challenge to efforts to describe biological phenomena using a classical (Kolmogorovian) mathematical framework. We develop a quantum framework for exaptation with examples from both biological and cultural evolution. The state of a trait is written as a linear superposition of a set of \emph{basis states}, or possible forms the trait could evolve into, in a complex Hilbert space. These basis states are represented by mutually orthogonal unit vectors, each weighted by an amplitude term. The choice of possible forms (basis states) depends on the adaptive function of interest (\emph{e.g.}, ability to metabolize lactose or thermoregulate), which plays the role of the \emph{observable}. Observables are represented by self-adjoint operators on the Hilbert space. The possible forms (basis states) corresponding to this adaptive function (observable) are called \emph{eigenstates}. 
The framework incorporates key features of exaptation: potentiality, contextuality, nonseparability, and emergence of new features. However, since it requires that one enumerate all possible contexts, its predictive value is limited, consistent with the assertion that there exists no biological equivalent to ``laws of motion" by which we can predict the evolution of the biosphere. 
\end{abstract}
\end{frontmatter}

\section{Introduction}

\noindent 
Representations of the underlying laws of how new traits, species, and cultural artifacts come into existence are outside the scope of current evolutionary biology. It is increasingly recognized that to capture the causal dynamics of biological systems there is a need for a general theory of biocomputation and novel mathematical formalisms capable of incorporating the multiple interacting facets of complex living systems \citep{Simeonovetal11}.  One central and fascinating feature of evolutionary change in need of a formal theoretical framework is \emph{exaptation}: the retooling or co-option of existing organs, appendages, or other evolved structures for new functions, possibly after further modification.  A model of exaptation must incorporate the notion of \emph{potentiality}: every biological change not only has direct implications for fitness and so forth, but it both enables and constrains potential future changes. The notion of potentiality incorporates both the `adjacent possible' \citep{Kauffman08}, those states that are directly achievable given a certain initial state, and the `nonadjacent possible', those states that are remotely achievable given a certain initial state. Exaptation occurs when selective pressure causes this  potentiality to be exploited.

This paper sketches a formal framework for modeling exaptation in which the notion of potentiality plays a central role. First we discuss the notion of exaptation as it applies at different levels of biological systems, and how it also applies in cultural evolution. We then discuss the challenges of developing a formal model of exaptation, and in particular why we believe a classical probabilistic framework is insufficient. Then we introduce the basic elements of a quantum-inspired formal framework for explaining evolutionary change, and show how the essential elements of both cultural and biological exaptation can be represented using vectors in a complex Hilbert space.  The paper concludes with a discussion of the strengths and limitations of the approach, and suggests empirical criteria for testing the predictions of the quantum probabilistic model.

A physics-inspired model may seem like a strange move given the position (as stated in the introduction to this special issue) that biology is `broken' in large part because it has adopted the abstractions of Newtonian physics \citep{Simeonovetal11}. However, the formalism we use does not come from Newtonian physics; indeed it arose through recognition of the limitations of Newtonian physics.  Moreover, many if not most branches of mathematics grew out of specific applications to physical systems that were subsequently generalized. It is perhaps unfortunate that the term `quantum' is associated with this approach to probability, given that it uses a generalization of quantum mechanics that has little to do with the quantum \emph{per se}. We are not attempting to argue that exaptation directly depends on the physics of quantum mechanics \emph{per se}, but rather using an abstract generalization of the quantum formalism. 

\section{Biological Exaptation}
\label{sec:Biology}
\noindent The term \emph{exaptation} was coined by \cite{GouldVrba} to denote what Darwin referred to as \emph{preadaptation}.\footnote{The terms \emph{exaptation}, \emph{preadaptation} and \emph{co-option} are often used interchangeably.} Both are used to refer to a situation in which a biological character serves a `current use' (possibly after some modification) that is different from the function it was originally adapted for. Like other kinds of evolutionary change, exaptation can be observed across all levels of biological organization.  It is appropriate to speak of exapted genes, tissue, organs, limbs and/or behaviors.  In this section we briefly review ways that exaptation is studied and attested in biology, so as to motivate the theory presented in later sections.

Classic examples come from morphology: ``A well-developed tail," speculated \cite{Darwin}, ``having been formed in an aquatic animal, it might subsequently come to be worked in for all sorts of purposes, -- as a fly-flapper, an organ of prehension, or as an aid in turning."  Likewise, the recent discovery of feathered non-avian theropods indicates that feathers evolved before flight -- likely for thermoregulation -- and were later co-opted \citep{PrumBrush}.

Another morphological example of exaptation, which we will return to below to illustrate our framework, is the swim bladder: a sac found in some fish, partially filled with water and partially filled with air, that adjusts neutral buoyancy in the water column.  Swim bladders are believed to have evolved by exaptation from primitive lungs \citep{DanielsEtAl04, PerryEtAl01}. According to this hypothesis, one or more populations of lungfish developed a propensity to collect water in their lungs.  Now there was a sac partially filled with air, partially with water, and so poised to evolve into a swim bladder (after some further modification). A new function thus arose in the biosphere: neutral buoyancy in the water column. The swim bladder in turn affected the further evolution of the biosphere by influencing the formation of new species, proteins, other molecules, niches, and so forth. 

Adaption and exaptation of characters such as these is typically tested for by constructing the maximum parsimony phylogenetic tree for the clade under consideration, and thus inferring the most likely evolutionary history and selective environment that operated on the trait.  Recently, a variety of more powerful statistical techniques have become increasingly popular due to their ability to (1) handle more uncertainty about phylogenetic history than parsimony, (2) incorporate population genetic models and data from other points on the evolutionary tree, and (3) quantitatively discriminate between competing hypotheses (e.g.\ \cite{BaumDonoghue, MacLeod, Martins, Smith}). As with any hypothesis concerning adaptive versus non-adaptive evolutionary influence, however, the typical paucity of historical data in many cases makes it difficult to devise an effective test for the occurrence of exaptation.

These problems are attenuated in the molecular realm, where exaptation is empirically well-attested as a major player in genetic change.  Spontaneous emergence of novel functional proteins of more than a few dozen amino acid residues \emph{de novo} is rare, since the probability of multiple useful mutations simultaneously arising decreases exponentially with a gene's complexity \citep{Patthy}.  The primary source of new proteins and regulatory elements is the duplication, tweaking, combination and subsequent repurposing of previously existing genes and small functional units \citep{BaileyEichler, Eichler, LongEtAl03, TaylorRaes}. For instance, phylogenetic analysis of sequential, structural, and functional relationships amongst proteins consistently and explicitly shows the tendency of new proteins to evolve out of copies of old ones, often diverging in their subsequent function. As a result, approximately three quarters of the many millions of gene sequences across all known species -- with their impressive retinue of diverse functions -- fit into on the order of just ten thousand families of related proteins \citep{FinnEtAl}. In addition to simple duplication and divergence of whole genes, many proteins consist of one or more \emph{domains}, that serve as sub-modules that are shuffled and recombined to create novel proteins \citep{ChothiaEtAl}. Up to 90\% of protein domains found in eukaryotes have been reused in more than one gene \citep{OrengoThornton}. On a higher level, whole networks of proteins can be reused for new adaptive purposes.  Most famous is the so-called `developmental toolkit,' a set of homologous genes -- strongly conserved across metazoans -- that is reused to control the ontogenesis of a wide diversity of morphologies \citep{CarrollGrenierWeatherbee}.

These discoveries give a high-resolution picture of a molecular evolutionary process that achieves novel functions primarily via recombination or `tinkering' with previously existing material \citep{Jacob, BarabasiOltvai}. 
\cite{GerhartKirschner07} argue that the possible phenotypic variation -- the `adjacent possible' in Kauffman's terms -- is both facilitated and constrained by pre-existing material:
\begin{quotation}
\noindent The burden of creativity in evolution, down to minute details, does not rest on selection alone. Through its ancient repertoire of core processes, the current phenotype of the animal determines the kind, amount, and viability of phenotypic variation the animal can produce in response to regulatory change. Thanks to the nature of the processes, the range of possible anatomical and physiological variations is enormous, and many are likely nonlethal, in part simply because the processes have been providing ``useful" function since pre-Cambrian times.
\end{quotation}
According to this exaptation-centric, `toolkit' view, complex function can only evolve if pre-existing genetic material facilitates it. Adaptive landscapes are seen as plagued with local optima, and only rarely does a particular genetic configuration confer the possibility of a ridge in the adaptive landscape from a local optimum to a point with higher fitness. This can be contrasted with the common \emph{adaptationist} perspective, which minimizes the importance of genetic and developmental limitations and assumes that species can reach optimal adaptations with relative ease \citep{OrzackSober}.

Adaptationist assumptions, though a fiction, allow elegant mathematical approaches to evolutionary game theory, playing the same role as the `rational actor' assumption in economics \citep{Gintis, Nowak}. Just as it proves a tricky matter to model economic decision making without this assumption (i.e., to introduce `bounded rationality'), it is by no means clear how to transmute the highly constrained `toolkit' understanding of evolution into a predictive or quantitative theory of exaptation's role in genetic and phenotypic change. This arises in part because it is not clear how to enumerate the possible states a character may evolve into or the selective environments that may direct it. Lest ``fools rush in where angels fear to tread," then, we must proceed incrementally in attempting to define a causal relationship between selective pressure and form.  This is precisely the goal of evolutionary developmental biology, but we are a long ways off from the requisite understanding of how genetic change entails phenotypic variation in nature, and how selection operates on different levels of the organism.

Closed-form analytical tools not forthcoming, a computational approach may prove fruitful.  Several recent computer models have reproduced exaptation in simple evolutionary problems, often using the evolution of Boolean functions under varying selective environments or similar toy problems \citep{ArthurPolak, Fentress, Graham, Lenski, Mouret, Mouret2, Oppacher, ParterKashtan, Skolicki}.  In artificial intelligence and cognitive science, exaptation finds 
analogues in the fields of `transfer learning' \citep{PanYang, TaylorStone, TorreyShavlik}, `shaping' or `scaffholding' in robotics \citep{Bongard, DorigoColombetti}, and the computational modeling of analogy-making \citep{Gentner98}, all of which concern the reuse of previously-learned material to aid in solving new problems.  While the methods and results found in these systems can provide a 
starting point for the study of exaptation in its interdisciplinary instantiations, in general they only succeed in producing exaptive events in contrived scenarios.  Since the potential for exaptation is inherently difficult to predict, these technologies rarely prove efficacious in the real-world applications of interest to computer scientists.

We hold that a satisfactory or useful understanding of exaptation and the reuse of information as a deeply ``facilitative" evolutionary tool, in the sense of \cite{GerhartKirschner07}, is unlikely to be approachable by either predictive theory or simulation until an artificial system is developed which displays a breed of immensity, diversity and/or open-ended development that mimics natural systems.  A high-level theoretical approach to describing the potential for exaptation, in the meantime, may be able to partially fill and motivate new approaches to this hole in our understanding of evolutionary processes.  The rest of this paper is an effort to make progress on this latter goal.

\section{Cultural Exaptation}

\noindent Cultural change not only accumulates over time, but it adapts, diversifies, becomes increasingly complex, and exhibits phenomena observed in biological evolution such as niches, drift, epistasis, and punctuated equilibrium \citep{Bentleyetal04, Durham91, Gabora01, Gabora95}. Processes of both biological and cultural evolution tend to gravitate toward a balance between differentiation (or divergence) and synthesis (or convergence) of different forms. Moreover, like biological evolution, culture is open-ended; there is no limit to the variety of new forms it can give rise to. Thus many have suggested that culture is a second evolutionary process which, though it piggybacks on the first, cannot be reduced to biology  \citep{Arthur, BoydRicherson85, BoydRicherson05, Cavalli-SforzaFeldman81, Gabora96, Gabora08, Gabora12, JablonkaLamb05, MesoudiWhitenLaland06}. 

Cultural evolution depends on the characteristically human capacity to combine concepts in new ways or redefine one concept by re-examining it in the context of another concept. It has been proposed that this is what ushered forth what  \cite{Mithen96} refers to as the `big bang of human creativity' characterized by the ``birth of art, science, and religion" in the Middle-Upper Paleolithic \citep{Gabora03}. An example of redefining one concept by re-examining it in the context of another is the tire swing. The tire swing came into existence when someone re-conceived of a tire as an object that could form the part of a swing that one sits on. It is this repurposing of an object designed for one use for use in another context that we refer to as  \emph{cultural exaptation}. Much as the current structural and material properties of an organ or appendage constrain possible re-uses of it, the current structural and material constraints on a cultural artifact (or language, or art form...) constrain possible re-uses of it.

\section{Challenges for a Formal Model of Exaptation}

\noindent Exaptation poses several challenge for those who wish to develop a formal model of it. 
One challenge is the \emph{highly contextual} nature of exaptation: change from one form to another reflects selective pressures offered by an ever-changing adaptive landscape, of which we lack complete knowledge. 
Another challenge is that exaptation entails \emph{emergence of novelty} because the co-opted body part, organ, or trait carries out a new adaptive function (\emph{e.g.,} the ability to adjust neutral buoyancy in the water column), and indeed the new form may be so different from its predecessor that it is thought of and referred to by a new name (\emph{e.g.,} swim bladder).
Moreover, this emergent novelty in exaptation may be \emph{non-compositional} because the whole (\emph{e.g.,} swim bladder) is not a simple function (such as logical AND or OR) of its constituents (\emph{e.g.,} an air sac and a water-dwelling organism). 
It also exhibits \emph{nonseparability} because the context (\emph{e.g.,} \underline{water dwelling}) becomes an inextricable part of the air sac in its new form (swim bladder), with its new function (adjusting neutral buoyancy). In other words, any modification or manipulation of the air sac simultaneously affects the object of selective pressure induced by \underline{water dwelling}, for these two are one and the same organ, the swim bladder. 

These challenges -- contextuality, emergence of novelty, non-compositionality, and nonseparability -- are not unique to biology; they also arise in psychology. 
For the last quarter century they have plagued psychologists' efforts to model how new meanings emerge when people combine concepts and words into larger semantic units such as conjunctions, phrases, or sentences. A compound concept's constituents are not just conjointly activated but bound together in a context-specific manner that takes relational structure into account \citep{GagneSpalding}. Copious empirical data shows that people use conjunctions and disjunctions of concepts in ways that violate the rules of classical (fuzzy) logic; \emph{i.e.}, concepts interact in ways that are non-compositional \citep{Hampton88, AertsAertsGabora09, KittoEtAl11, OshersonSmith81}. This is true both with respect to properties (\emph{e.g.}, although people do not rate `talks' as a characteristic property of \texttt{PET} or \texttt{BIRD}, they rate it as characteristic of \texttt{PET BIRD}), and exemplar typicalities (\emph{e.g.}, although people do not rate `guppy' as a typical \texttt{PET}, nor a typical \texttt{FISH}, they rate it as a highly typical \texttt{PET FISH}).  In other words, if something is an instance of \texttt{PET FISH}, it is not possible to perturb the \texttt{PET} without simultaneously affecting the  \texttt{FISH}, for these two are one and the same, a \texttt{PET FISH}. This non-compositional emergence of new properties in new contexts has made concepts particularly resistant to mathematical description. 

\section{Generalized Quantum Models}
\label{sec:GeneralizedQuantum}
The approach taken in this paper to model exaptation builds on early efforts toward a cross-disciplinary framework for evolution, which focused on distinguishing processes according to the degree of non-determinism they entail, and the extent to which they are sensitive to, internalize, and depend upon particular contexts \citep{GaboraAerts05b, GaboraAerts08}. 
We define \emph{context} to refer to anything that is not part of the entity of interest (such as a particular organ or appendage) that needs to be included in our model (such as an aspect of the environment that exerts selective pressure). Processes were modeled as re-iterated \emph{context-dependent actualization of potential}, or CAP: an entity has potential to change various ways, and how it \emph{does} change depends on the contexts it interacts with. Potentiality and contextuality can be viewed as flip sides of the same coin in the sense that the interaction between entity and context actualizes some of the potentiality of that entity. 

If an entity in a state \(p(t_i)\) at time \(t_i\) under the influence of a context \(e(t_i)\) could potentially change to more than one state at \(t_{i+1}\) then we may say that the process is \emph{nondeterministic}. There are different ways that nondeterminism can arise, and accordingly, different kinds of probability spaces. Conditional probability -- the most straightforward way to build a probabilistic model -- involves one equivalent probability space, and it can be modeled using a classical Kolmogorovian probability model. Because a classical Kolmogorovian probability model uses the same probability space for all contexts, it is of limited use in modeling situations that exhibit extreme contextuality, and the sort of noncompositional, nonseparable, emergent novelty it can give rise to.  

The first field to deal with potentiality and contextuality in a rigorous way is quantum mechanics. 
It has been shown that there exist macroscopic contextual systems that display no quantum mechanical effects in a physical sense wherein uncertainty arises due to the measurement process itself, and that the formalism of quantum mechanics may provide a probability calculus to model such systems \citep{Aerts83}. Generalizations of it have been extensively applied to areas such as information retrieval \citep{BruzaCole05, Grover97, KittoEtAl11, NielsenChuang00}, economics \citep{Baaquie04}, and psychology -- in particular concept combination \citep{Aerts09, Aertsetal12, AertsGaboraSozzo, Gabora01, GaboraAerts02, GaboraAerts05b, GaboraAerts09} and decision making \citep{AertsAerts95, BusemeyerWangTownsend06A, BusemeyerWangTownsend06B, BusemeyerPothosFrancoTrueblood11, PothosBusemeyer12}.  The rationale for applying quantum formalisms to macroscopic systems is covered elsewhere \citep{Aertsetal00, BruzaBusemeyerGabora09}, including in this volume \citep{KittoKortschak}.

The advantage of a quantum model over the classical one is that it uses variables and spaces that are defined specifically with respect to a particular context, and it uses amplitudes, which though directly related to probabilities, can exhibit \emph{interference} when used in a complex Hilbert space (which we will define in a moment).  A related key feature that can be modeled using complex Hilbert space, \emph{entanglement}, was specifically conceived to deal with situations of nonseparability wherein entities interact in such a way that each entity is described by the same quantum mechanical description. Entangled states may be said to exhibit \emph{non-compositionality} because they may exhibit emergent properties not inherited from the constituent parts.

The features thus identified that make quantum models applicable to concept combination and other areas match the challenges for the formal modeling of exaptation which we delineated above. In what follows, we provide a general scheme for modeling exaptation using the quantum approach.  

In quantum mechanics, the \emph{state} \(| \Psi \rangle\) of an entity is written as a linear superposition of a set of \emph{basis states} \(\{| \phi_i \rangle\}\) of a \emph{Hilbert space} \(\mathcal{H}\) which is a real or complex vector space. Each complex number coefficient of the linear superposition, referred to as the \emph{amplitude} and denoted \(a_i\), represents the contribution of each component state \(|\phi_i\rangle\) to the state \(|\Psi\rangle\). Hence \(| \Psi \rangle = \sum_ia_i | \phi_i \rangle\). The square of the absolute value of the amplitude equals the probability of its component basis state with respect to the global state. The choice of basis states is determined by the \emph{observable}, \(o_i \in \mathcal{O}\), to be measured. 
The basis states corresponding to an observable are called \emph{eigenstates}. 
Observables introduce particular symmetry transformations and are represented by self-adjoint operators that define subspaces on the Hilbert space. 
The lowest energy state of the entity is referred to as the \emph{ground state}. 
Upon \emph{measurement}, the state of the entity \emph{collapses} out of the ground state and it is projected onto one of the eigenstates. 

Consider two entities $A$ and $B$ with Hilbert spaces $\mathcal{H_A}$ and $\mathcal{H_B}$, where is the amplitude associated with the first is \(a_i^A\) and is the amplitude associated with the second is \(a_j^B\). 
The Hilbert space of the composite of these entities is given by the tensor product  $\mathcal{H_A} \otimes \mathcal{H_B}$.
We may define a basis  ${| i \rangle_A}$ for $\mathcal{H_A}$ and a basis  ${| j \rangle_B}$ for $\mathcal{H_B}$.  The most general state in $\mathcal{H_A} \otimes \mathcal{H_B}$ has the form 
\begin{equation} 
| \Psi \rangle_{AB} = {\sum}_{i,j} a_{ij} | i \rangle_A \otimes | j \rangle_B\
\end{equation}
This state is separable if $a_{ij}=a_i^Aa_j^B$.
It is inseparable, and therefore an entangled state, if  $a_{ij}\neq a_i^Aa_j^B$.

In some applications the procedure for describing entanglement is more complicated than what is described here. For example, it has been argued that the quantum field theory procedure, which uses Fock space to describe multiple entities, gives a kind of internal structure that is superior than the tensor product for modeling concept combination (Aerts, 2007, 2009). Fock space is the direct sum of tensor products of Hilbert spaces, so it is also a Hilbert space. For simplicity this initial application to exaptation will omit such refinements, but such a move may become necessary in further developments of the model. 

\section{Quantum Model of Exaptation}
\label{sec:QuantumExaptation}

\noindent The quantum approach is applied to exaptation as follows.
The set of possible states of a particular trait is given by $\Sigma$.\footnote{As noted in section~\ref{sec:Biology}, it is not clear how to enumerate this set in practice, much less their probabilities.  This is the fundamental difficulty in attempting to predict evolutionary change, which we return to in section~\ref{sec:Testing}.} The current state of the trait  $| p \rangle$ is written as a linear superposition of basis states in a complex Hilbert space \(\mathcal{H}\) each of which represent a possible form that the trait could evolve into.
The amplitude term associated with a basis state represented by a complex number coefficient \(a_i\) gives a measure of how likely a given evolutionary change of state is. 
The basis states represent possible forms of the organ or appendage of interest. States are represented by unit vectors, and all vectors of a decomposition have unit length, are mutually orthogonal, and generate the whole vector space, thus \(\sum_i|a_i|^2 = 1\).
In generalizations wherein the quantum formalism is applied to other domains, the self-adjoint operators are used to define context-specific subspaces. The particular adaptive function of interest (\emph{e.g.}, ability to metabolize lactose, or to thermoregulate) plays the role of a measurement in physics, causing the state of the trait to collapse to one of its eigenstates.  The role of the observable is played by the detectable changes to the trait in question. Thus we model change in the trait's function under evolutionary forces (including natural selection and drift) by collapse to a new state. Rarely is a single change of state involved -- a noticeable change of state may consist of a sequence of barely detectable micro-changes -- but for simplicity we focus on one change of state. Clearly, since exaptation may involve some modification of the biological character as it adopts a new function, this process is not genuinely instantaneous.  We make the simplifying assumption that this process of change is transient, and that the change can ultimately be interpreted as a discrete transition to a new local optimum.

The environment or \emph{context}  in which the change of state is taking place is for simplicity represented \(c_i \in \mathcal{C}\). Note that what is considered the trait with respect to one evolutionary change of state process (such as the change of state of an organ in one particular species), may be considered the context with respect to another evolutionary change of state process (such as the change of state of organs in predators or prey of that species).

Each possible form of a trait represented by a particular basis state can be broken down into a set \(f_i \in \mathcal{F}\) of features (or properties), which may be weighted according to their relevance with respect to the current context.
The \emph{weight} (or renormalized applicability) of a certain property given a specific state of the trait $| p \rangle$ and a specific context $c_i \in \mathcal{C}$ is given by $\nu$. 
For example, $\nu (p, f_1)$ is the weight of feature ${f_i}$ for state $p$. Thus $\nu$ is a function from the set $\Sigma \times {\cal F}$ to the interval $[0, 1]$. We write:
\begin{eqnarray}\
\nu: \Sigma 
\times {\cal F} &\rightarrow& [0, 1] \\
(p, f_i) &\mapsto& \nu(p, f_i) 
\nonumber
\end{eqnarray}

A function $\mu$ describes the transition probability from one state to another under the influence of a particular context. For example, $\mu$\textit{(q, e, p)} is the probability that state \textit{p} under the influence of context \textit{e} changes to state \textit{q}. Mathematically, $\mu$ is a function from the set $\Sigma \times \mathcal{C} \times \Sigma $
to the interval $[0, 1]$, where $\mu(q, e, p)$ is the probability that state $p$ under the influence of context $e$ changes to state $q$. We write: 
\begin{eqnarray}
\mu: \Sigma \times \mathcal{C} \times \Sigma 
&\rightarrow& [0, 1]  \\
(q, e, p) &\mapsto& \mu(q, e, p) 
\nonumber
\end{eqnarray}
Thus our quantum model of exaptation consists of the 3-tuple $(\Sigma, \mathcal{C}, \mathcal{F})$, and the functions $\nu$ and $\mu$. 
We will now present two examples of the quantum model of exaptation. First the approach will be applied to an example of cultural exaptation, since cultural exaptation arises more directly from the quantum-inspired models of concept combination discussed previously. Next the approach will be applied to an example of biological exaptation.

\subsection{Modeling Cultural Exaptation}

\noindent Cultural evolution depends on the capacity for individuals to combine concepts in new ways or redefine one concept by re-examining it in the context of another concept. Let us see how a small step in cultural evolution could take place through the reconceptualization of a particular concept, the concept \texttt{TIRE}.

The state of \texttt{TIRE}, represented by vector \(|p\rangle\) of length equal to 1, is a linear superposition of basis states in a complex Hilbert space \(\mathcal{H}\) which represent possible states (new types or versions) of this concept, such as \texttt{SNOW TIRE} or \texttt{BIKE TIRE}. In the context  \underline{winter}, \texttt{TIRE} might collapse to \texttt{SNOW TIRE} and in the context \underline{bicycle} it might collapse to \texttt{BIKE TIRE}. 

\begin{figure} \label{fig:superposition}
\begin{center}
	\includegraphics[scale=.25]{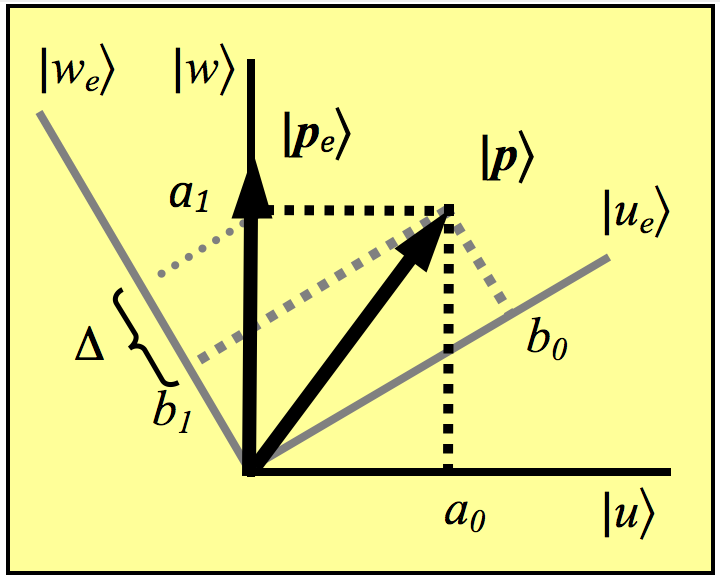}
\end{center}
\caption
{ 
Graphical depiction of a vector $| p \rangle$ representing particular state of \texttt{TIRE}, specifically, a state in which the tread is worn away. 
In the default context, the state of tire is more likely to collapse to the projection vector $| w \rangle$ which represents wasteful than than to its orthogonal projection vector $| u \rangle$ which represents useful. 
This can be seen by the fact that subspace $a_0$ is smaller than subspace $a_1$. Under the influence of the context \underline{playground equipment}, the opposite is the case, as shown by the fact that $b_0$ is larger than  $b_1$. 
Also shown is the projection vector after renormalization.
}
\end{figure}

Suppose that the initial conception of \texttt{TIRE} is a superposition of only two possibilities (Figure~\ref{fig:superposition}). The possibility that the tire has sufficient tread to be \emph{useful} is denoted by the unit vector \(|u\rangle\). The possibility that it should be discarded as \emph{waste} is denoted by the unit vector \(|w\rangle\). Their relationship is given by the equation 
\begin{equation} \label{}
|p\rangle = a_0|u\rangle + a_1|w\rangle,
\end{equation}
where \(a_0\) and \(a_1\) are the amplitudes of \(|u\rangle\) and \(|w\rangle\) respectively in the mind of a particular individual. In a different individual, who has had different experiences, and has a slightly different way of thinking things through,  \(a_0\) and \(a_1\) might be different, as epitomized in the saying ``one person's trash is another person's treasure". indeed as we will see, this can even be the case in the mind of the same individual thinking of the same thing from a new perspective.
States are represented by unit vectors and all vectors of a decomposition such as \(|u\rangle\) and \(|w\rangle\) have unit length, are mutually orthogonal and generate the whole vector space; thus \(|a_0|^2 + |a_1|^2 = 1\). 

If a tire is useful only for transportation, denoted \(|t\rangle\) then, \(|u\rangle = |t\rangle\). In the mind of the individual thinking about tires, the conception of \texttt{TIRE} changes when activation of the set \(\mathcal{L}\) of properties of \texttt{TIRE}, \emph{e.g.} the property `weather resistant' denoted $f_1$, spreads to other concepts for which these properties are relevant. Contexts such as \underline{playground equipment} that share properties with \texttt{TIRE} become candidate members of the set \(\mathcal{C}\) of relevant contexts for \texttt{TIRE}. Given the context \underline{playground equipment} denoted \(e\), some possible (however unlikely) states of a tire are to use it as a swing or to use it as a slide. We denote \texttt{SWING} and \texttt{SLIDE} as \(|s\rangle\) and \(|l\rangle\), respectively. 
It seems reasonable to designate the default context for these two states as \underline{playground equipment}; thus these states in this context are indicated  \(|s_e\rangle\) and \(|l_e\rangle\). The restructured conception of \texttt{TIRE} in the context of \underline{playground equipment}, denoted \(|p_e\rangle\), is given by 
\begin{equation} \label{2}
|p_e\rangle\ = b_0|u_e\rangle + b_1|w_e\rangle\,
\end{equation}
where 
\begin{equation} \label{3}
|u_e\rangle = b_2|t_e\rangle + b_3|t_es_e\rangle + b_4|t_el_e\rangle,
\end{equation}
and where \(|t_e\rangle\) represents the possibility that in the context \underline{playground equipment} the worn-out tire somehow manages to function as a tire, \(|t_es_e\rangle\) represents the possibility that in this context a tire functions as a swing, and \(|t_el_e\rangle\) stands for the possibility that in this context a tire functions as a slide. The overall probability that the tire is conceived of as useful has increased since \(|b_0|\) consists of the possibility of a tire being used not just as a tire, but as a swing or slide. 

Consider the set of strongly weighted properties of \texttt{SLIDE}, such as `long' denoted $f_2$ and `has steps', denoted $f_3$. Because `long' and `has steps' are not properties of \texttt{TIRE}, $\nu(p, f_2) << \nu(l, f_2)$, and similarly $\nu(p, f_3) << \nu(l, f_3)$. Therefore, \(|b_4|\) is small. This is not the case for \texttt{SWING}. Consider the property `has surface to sit on', denoted $f_4$. Since one could sit on a tire, $\nu(p, f_4) \approx \nu(s, f_4)$. 
Therefore, \(|b_3|\) is large. Thus $\mu(s, e, p) >> \mu(l, e, p)$. In the context \underline{playground equipment}, the concept \texttt{TIRE} has a high probability of collapsing to \texttt{TIRE SWING}.
A tire swing has the emergent property of a `bottomless' seat, i.e., a hole in the centre of where one sits. The quantum formalism was developed in part specifically to model the emergence of new properties using the notion of entanglement. 
The formalism can describe \texttt{TIRE SWING} as an entangled state of the concepts \texttt{TIRE} and \texttt{SWING}. If this collapse takes place, \texttt{TIRE SWING} is thereafter a new state of both concepts \texttt{TIRE} and \texttt{SWING}. Entanglement introduces interference of a quantum nature, and hence the amplitudes are complex numbers \citep{Aerts09}. 

This example shows that a formal approach to concept interactions that has been previously shown to be consistent with human data \citep{Aerts09, Aertsetal12} can model the restructuring of a concept (\emph{e.g.}, \texttt{TIRE}) under the influence of a new context (\emph{e.g.}, \underline{playground equipment}). The quantum approach is necessary to model an entangled state of \texttt{TIRE} and \texttt{SWING}. 

\subsection{Modeling Biological Exaptation}
\noindent 
Now let us use the same approach to model a biological example of exaptation: the swim bladder. Like \texttt{TIRE}, the trait \texttt{AIR SAC} can take different forms, \emph{e.g.,} one state of \texttt{AIR SAC} is \texttt{LUNG}, and another is \texttt{SWIM BLADDER}. Each state consists of a set \(\mathcal{F}\) of features or properties. For example, some possible properties of \texttt{AIR SAC} are \emph{thin-walled}, \emph{capillary-rich}, and \emph{air-filled}. The applicability or \emph{weight} \emph{w} of a certain property depends on the specific state and context. For example, given the state \texttt{LUNG} of \texttt{AIR SAC} and the context \underline{air-breathing}, the weight of \emph{air-filled} would be high. 

The context \underline{water-dwelling}, denoted \(e\), is one of several members of the set \(\mathcal{C}\) of contexts that bias the evolution of traits such as \texttt{AIR SAC}. Other such contexts might be \underline{high predation} or \underline{low oxygen}. The  \emph{observable}, \(a \in \mathcal{O}\), to be measured is the ability to adjust neutral buoyancy in the water column. The amplitude term represented by a complex number coefficient \(a_i\)
of the linear superposition gives the probability of changing from \texttt{AIR SAC} to \texttt{SWIM BLADDER} under the context \underline{water-dwelling}. The state of \texttt{AIR SAC}, represented by vector \(|p\rangle\) of length equal to 1, can be represented as a superposition of possible states. The possibility that it is useful is denoted by unit vector, \(|u\rangle\). The possibility that it wastes away and becomes merely a \texttt{VESTIGIAL LUNG} is denoted by unit vector \(|w\rangle\). Their relationship can be described by the equation 
\begin{equation} \label{}
|p\rangle = a_0|u\rangle + a_1|w\rangle,
\end{equation}
where \(a_0\) and \(a_1\) are the amplitudes of \(|u\rangle\) and \(|w\rangle\) respectively. 

Given an organ with a set of particular properties, there are some changes of state it could undergo that would be useful with respect to the context \underline{water-dwelling}. Alternatively, given another set of properties, an organ might be in an eigenstate with respect to that context. The change of state of \texttt{AIR SAC} in the context of \underline{water-dwelling}, denoted \(|p_e\rangle\), is given by 
\begin{equation} \label{}
|p_e\rangle = b_0|u_e\rangle + b_1|w_e\rangle.
\end{equation}

We may consider different possible ways in which a body part can be useful with respect to the context \underline{water-dwelling}. One might be that it aids digestion of plankton. Another might be that it makes the body streamlined. Thus
\begin{equation} \label{}
|u_e\rangle = b_2|p_ed_e\rangle + b_3|p_es_e\rangle + b_4|p_en_e\rangle + b_5|p_el_e\rangle
\end{equation}
where 
\(|p_ed_e\rangle\) stands for the possibility that in the context \underline{water-dwelling} this organ functions to \texttt{AID DIGESTION OF PLANKTON}, 
\(|p_es_e\rangle\) stands for the possibility that it functions to \texttt{MAKE BODY STREAMLINED}, and
 \(|p_en_e\rangle\) stands for the possibility that it functions to \texttt{ADJUST NEUTRAL BUOYANCY}, and for completeness we add 
 \(|p_el_e\rangle\) which stands for the unlikely possibility that it functions as a \texttt{LUNG}.

We know that $b_2$ and $b_3$ are small because the properties that could assist these functions overlap not at all with the properties of \texttt{LUNG}. Thus, there are no `adjacent possible' changes of state for this body part that will allow it to assist with these tasks. However, accidental entry of water into the \texttt{LUNG} takes it closer to the capacity to \texttt{ADJUST NEUTRAL BUOYANCY}, so  $b_4$ is large. 
Thus $\mu(n, e, p) >> \mu(d, e, p)$ and also $\mu(n, e, p) >> \mu(s, e, p)$. 
Therefore, in the context \underline{water-dwelling}, the trait of possessing an \texttt{AIR SAC} has a high probability of collapsing to \texttt{SWIM BLADDER}, which can be modeled as an entangled state of \texttt{AIR SAC} and \texttt{ADJUST NEUTRAL BUOYANCY} arising in the context \underline{water-dwelling}. Once again since the property of being able to adjust neutral buoyancy is an emergent property that was not present prior to the merging of concept and context, this is modeled as a state of entanglement, which requires complex numbers. Thus it is possible to use the quantum approach to formally model the restructuring of biological information through exaptation.

\section{Testing the Theory}
\label{sec:Testing}
\noindent The quantum framework for exaptation is admittedly speculative. The reader might wonder how it could be tested and how feasible the procedure is as a predictive tool in biology or as a novelty-generating tool the cultural realm. 
\cite{Longoetal12} claim that with respect to an evolving entity, it is not possible to define `random' or `equiprobable', or even to know the sample space. The possible uses of an entity such as an air sac or a screwdriver are indefinite in number, and un-orderable, so there is no procedure or algorithm that can list them all. Even if we limit ourselves to a single use of a screwdriver -- to open paint cans -- the number of objects or processes that can be used separately or together to open a paint can is indefinite and un-orderable.  In our terms, we cannot have a pre-stated, listed, basis set of all potential functions in a pre-stated Hilbert space with a pre-stated set of contexts. In short, the model (at least as it has been formulated here) requires that we enumerate possibilities up front, a procedure that one might well believe to be too laborious to make it useful as a predictive tool. 
They further argue that since because we cannot pre-state the ever-changing state space or phase space of an evolving entity, nor pre-state the context (the ``actual niche" in their terminology), we cannot, using the niche as boundary condition, integrate the equations of motion. Thus, in their view, evolution is not entailed by laws. 

However, the process need not be so laborious as it first appears. The `magic' of an associative memory, and indeed the trick to how the creative mind hits on useful connections without considering all possible connections, is that because associative memory is distributed and content-addressable, it can connect states and contexts \emph{on the basis of shared properties} \citep{Gabora10, GaboraRanjan13}. It is not necessary to consider all possible modifications of all possible organismal traits at all possible levels to make reasonable hypotheses concerning what exaptation events will be realized; one can jump immediately to probable exaptation events by looking exclusively at traits wherein the properties associated with the trait overlap substantially with the properties of a solution. 
For example, since both \texttt{AIR SAC} and \texttt{ADJUST NEUTRAL BUOYANCY} involve the property \emph{hold water}, a mechanism for adjusting neutral buoyancy is within the `adjacent possible' for \texttt{AIR SAC}.
A large genomic data base could in principle operate similarly using a neural network to make associations between states of organs and appendages, and selective forces or contexts that could alter their states, thus alleviating the need for brute force search through the space of all possibilities. Similarly, both \texttt{TIRE} and \texttt{SWING} involve the property \emph{can sit on}. We are developing an algorithm for development of waste recycling ideas that makes use of this principle. 

Moreover, the appropriateness of the model as an explanation of the causal forces involved can indeed be tested. To show how one might go about this, we first explain how the model has been tested with respect to concept combination, and then propose an analogous procedure for exaptation. 
As mentioned previously, people use conjunctions and disjunctions of concepts in ways that violate the rules of classical logic;  \emph{i.e.,} concepts interact in ways that are non-compositional. This has come to be known as the Pet Fish Problem, and the general phenomenon wherein the typicality of an exemplar for a conjunctively combined concept is greater than that for either of the constituent concepts has come to be called the
Guppy Effect, due to the well-known finding that although people do not rate \texttt{Guppy} as a typical \texttt{PET}, nor a typical \texttt{FISH}, they rate it as a highly typical \texttt{PET FISH} \citep{Osherson81}. When people rate the instance, or  \emph{exemplar}, as more typical of the conjunction that of its constituent concepts they are said to emph{overextend} it. 
In a well-known study of this phenomenon, \cite{Hampton88} collected data from human participants to determine the relative frequency of membership of specific exemplars of general categories or concepts, as well as of conjunctions of these concepts. For example, he asked participants whether an exemplar such as \texttt{Mint} is a member of \texttt{FOOD}, whether it is a member of \texttt{PLANT}, and whether it is a member of \texttt{FOOD AND PLANT}. 
For several of the items, participants assessed the examplar as more strongly a member of \texttt{FOOD AND PLANT} than of either of the two component concepts \texttt{FOOD} and \texttt{PLANT} alone.
For example, the relative frequency of membership for \texttt{Mint} was 0.87 for the concept \texttt{FOOD}, 0.81 for the concept \texttt{PLANT}, and 0.9 for the conjunction \texttt{FOOD AND PLANT}. 
It is difficult to conceive of a classical probability model that could encompass such a finding, and indeed it was subsequently shown using a geometric method that no such model exists. Specifically, it was proven that the membership weights $\mu(A), \mu(B)$ and $\mu(A\ {\rm and}\ B)$ can be represented within a classical probability model if and only if the following two requirements are satisfied (Aerts, 2009, theorem 3):

\begin{eqnarray} \label{mindeviation}
\mu(A\ {\rm and}\ B)-\min(\mu(A),\mu(B))=\Delta_c\le 0 \\ \label{kolmogorovianfactorconjunction}
0 \le k_c=1-\mu(A)-\mu(B)+\mu(A\ {\rm and}\ B)
\end{eqnarray}

where $\Delta_c$ is referred to as the `conjunction rule minimum deviation', and $k_c$, the `Kolmogorovian conjunction factor'. 
Applying this to the exemplar \texttt{Mint}, if $A$ is \texttt{FOOD} and $B$ is \texttt{PLANT}, inequality (\ref{mindeviation}) is violated, because plugging the above values into the first equation, we get  $\Delta_c = 0.9 - 0.81 = 0.09\not\le0$. Hence there is no classical probability representation for these data. It was, however, shown that the entire data set including the deviant items such as \texttt{Mint} can be modeled in a quantum framework \citep{Aerts09}.
Similar results were obtained with a data set involving exemplars such \texttt{Fridge} and the concepts \texttt{FURNITURE}, \texttt{HOUSEHOLD APPLIANCES}, and their conjunction \texttt{FURNITURE AND HOUSEHOLD APPLIANCES} \citep{Aertsetal12} and a data set involving exemplars such \texttt{Apple} and the concepts \texttt{FRUIT}, \texttt{VEGETABLE}, and their conjunction \texttt{FRUIT AND VEGETABLE} \citep{AertsGaboraSozzo}. 

It was shown that overextension of the conjunction could be modeled as an interference effect. Interference is a well-known phenomenon in quantum mechanics that was first demonstrated using the famous two-slit experiment, wherein the pattern that results when photons are detected on a screen after passing through two holes is different than what would be predicted by the results with only one hole or the other open. Interference effects cannot be described without complex numbers which necessitates a quantum formalism of the sort used here. The interference effect observed with concepts was viewed as, not a logical fallacy on the conjunction as would be suggested by a classical probability approach, but a signal that a new concept has emerged out of the two constituent concepts.

An analogous technique could be applied to the example of exaptation described above. Much as in the two slit experiment, different food items can be categorized as instances of \texttt{FRUIT}, instances of \texttt{VEGETABLE}, or instances of \texttt{FRUIT AND VEGETABLE}, different forms of an air sac qualify to different degrees as indicative of an animal that is \texttt{AIR-DWELLING}, an animal that is \texttt{WATER-DWELLING}, and an animal that is both \texttt{AIR-DWELLING AND WATER-DWELLING}. 
 \texttt{AIR-DWELLING} and \texttt{WATER-DWELLING} play the role of \texttt{FRUIT} and \texttt{VEGETABLE}, the equivalent of the two slits in the two-slit experiment. 
 \texttt{AIR-DWELLING AND WATER-DWELLING} plays the role of \texttt{FRUIT AND VEGETABLE}, the equivalent of having both slits open in the two-slit experiment. 
One shows photographs of various lungs, gills, swim bladders, and unusual or intermediate forms of these organs, to expert biologists. For each picture, they are asked whether it would be expected in an animal that is \texttt{AIR-DWELLING} an animal that is \texttt{WATER-DWELLING}, and an animal that is both \texttt{AIR-DWELLING AND WATER-DWELLING}. 
We then determine whether the data exhibit a deviation from what would be classically predicted and if this deviation can be modeled as using interference. A more sophisticated approach would be to tackle this computationally at the genomic level using a genetic algorithm in which the bitstrings are randomly mutated versions of the genomic region for a particular organ in a given species. We leave elaboration of these ideas for another paper. 

\section{Conclusions}

\noindent This paper may open up more questions than it answers, but we hope that this is not an entirely negative state of affairs. Our hope is that, at the very least, it provokes consideration of the interesting challenges involved in constructing a formal framework for exaptation, and at best, sketches a possible route forward. The fact that highly adaptive and highly maladaptive biological forms alike get actualized as living, physical organisms that compete for existence has enabled biology to get by with a kind of theorizing that all but ignores the notion of potentiality. 
Compare this to the novelty generating and pruning processes that underlie cultural evolution. If you want to entertain and explore the feasibility of different possible \emph{ideas}, you can imagine they exist, mentally simulate how they might work, and run through possible scenarios for how effective or interesting they might be if put into practice. You can `test drive' ideas before `taking them out on the road' so to speak.
It is perhaps because biological evolution does not explicitly incorporate this kind of `test drive' phase that biological theorizing has not been forced to seriously confront the notion of potentiality, and there has been relatively little to incorporate it into formal models of biological change. Psychologists have been forced to address the issue of potentiality to explain how concepts can `collapse' to different meanings in different contexts. 

In this paper we took what we believe to be a promising step toward incorporating the notion of potentiality into biological theory. We showed how exaptation has a cultural equivalent, and suggest a formal framework for modeling it. Although the example given involved exaptation at the level of the organ, the basic approach can be applied to exaptation events at the microbiological level. 
For simplicity we used a Hilbert space based model. Although we believe the approach in general is amenable to further developments, Hilbert space may turn out to be insufficient. 

We are still a long way from a satisfying account of how new traits, species, and cultural artifacts come into existence. However, the approach puts us on the path toward a formal framework that can accommodate exaptation, a process that plays a major role in the evolution of novel form, both biological and cultural. 

\section{Acknowledgments}
\noindent We thank Kirsty Kitto for comments on the manuscript. This work is supported by grants from the Natural Sciences and Engineering Research Council of Canada and the TEKES Foundation of Finland. \\

\bibliographystyle{model2-names}
\bibliography{pbmb2012}

\begin{thebibliography}{85}
\expandafter\ifx\csname natexlab\endcsname\relax\def\natexlab#1{#1}\fi
\expandafter\ifx\csname url\endcsname\relax
  \def\url#1{\texttt{#1}}\fi
\expandafter\ifx\csname urlprefix\endcsname\relax\def\urlprefix{URL }\fi
\providecommand{\eprint}[2][]{\url{#2}}
\providecommand{\bibinfo}[2]{#2}
\ifx\xfnm\relax \def\xfnm[#1]{\unskip,\space#1}\fi
\bibitem[{Aerts(1983)}]{Aerts83}
\bibinfo{author}{Aerts, D.}, \bibinfo{year}{1983}.
\newblock \bibinfo{title}{Classical theories and nonclassical theories as a
  special case of a more general theory}.
\newblock \bibinfo{journal}{Journal of Mathematical Physics}
  \bibinfo{volume}{24}, \bibinfo{pages}{2441--2454}.
\bibitem[{Aerts(2009)}]{Aerts09}
\bibinfo{author}{Aerts, D.}, \bibinfo{year}{2009}.
\newblock \bibinfo{title}{Quantum structure in cognition}.
\newblock \bibinfo{journal}{Journal of Mathematical Psychology}
  \bibinfo{volume}{53}, \bibinfo{pages}{314--348}.
\bibitem[{Aerts and Aerts(1995)}]{AertsAerts95}
\bibinfo{author}{Aerts, D.}, \bibinfo{author}{Aerts, S.}, \bibinfo{year}{1995}.
\newblock \bibinfo{title}{Applications of quantum statistics in psychological
  studies of decision processes}.
\newblock \bibinfo{journal}{Foundations of Science} \bibinfo{volume}{1},
  \bibinfo{pages}{85--97}.
\bibitem[{Aerts et~al.(2000)Aerts, Aerts, Broekaert and Gabora}]{Aertsetal00}
\bibinfo{author}{Aerts, D.}, \bibinfo{author}{Aerts, S.},
  \bibinfo{author}{Broekaert, J.}, \bibinfo{author}{Gabora, L.},
  \bibinfo{year}{2000}.
\newblock \bibinfo{title}{The violation of bell inequalities in the
  macroworld}.
\newblock \bibinfo{journal}{Foundations of Physics} \bibinfo{volume}{30},
  \bibinfo{pages}{1387--1414}.
\bibitem[{Aerts et~al.(2009)Aerts, Aerts and Gabora}]{AertsAertsGabora09}
\bibinfo{author}{Aerts, D.}, \bibinfo{author}{Aerts, S.},
  \bibinfo{author}{Gabora, L.}, \bibinfo{year}{2009}.
\newblock \bibinfo{title}{Experimental evidence for quantum structure in
  cognition}, in: \bibinfo{editor}{Bruza, P.}, \bibinfo{editor}{Lawless, W.},
  \bibinfo{editor}{van Rijsbergen, K.}, \bibinfo{editor}{Sofge, D.} (Eds.),
  \bibinfo{booktitle}{Proceedings of the Third International Conference on
  Quantum Interaction}, \bibinfo{address}{German Research Center for Artificial
  Intelligence, Saarbruken, Germany}. pp. \bibinfo{pages}{59--70}.
\bibitem[{Aerts et~al.(2012a)Aerts, Broekaert, Gabora and Veloz}]{Aertsetal12}
\bibinfo{author}{Aerts, D.}, \bibinfo{author}{Broekaert, J.},
  \bibinfo{author}{Gabora, L.}, \bibinfo{author}{Veloz, T.},
  \bibinfo{year}{2012}a.
\newblock \bibinfo{title}{The guppy effect as interference}, in:
  \bibinfo{booktitle}{Proceedings of the Sixth International Symposium on
  Quantum Interaction}, \bibinfo{address}{Paris}.
\bibitem[{Aerts et~al.(2012b)Aerts, Gabora and Sozzo}]{AertsGaboraSozzo}
\bibinfo{author}{Aerts, D.}, \bibinfo{author}{Gabora, L.},
  \bibinfo{author}{Sozzo, S.}, \bibinfo{year}{2012}b.
\newblock \bibinfo{title}{How concepts combine: A quantum theoretic modeling of
  human thought}.
\newblock \bibinfo{note}{ArXiv:1206.1069 [cs.AI]}.
\bibitem[{Arthur(2009)}]{Arthur}
\bibinfo{author}{Arthur, W.B.}, \bibinfo{year}{2009}.
\newblock \bibinfo{title}{The Nature of Technology: What It Is and How It
  Evolves}.
\newblock \bibinfo{publisher}{Free Press}.
\bibitem[{Arthur and Polak(2006)}]{ArthurPolak}
\bibinfo{author}{Arthur, W.B.}, \bibinfo{author}{Polak, W.},
  \bibinfo{year}{2006}.
\newblock \bibinfo{title}{The evolution of technology within a simple computer
  model}.
\newblock \bibinfo{journal}{Complexity} \bibinfo{volume}{11},
  \bibinfo{pages}{23--32}.
\bibitem[{Baaquie(2004)}]{Baaquie04}
\bibinfo{author}{Baaquie, B.E.}, \bibinfo{year}{2004}.
\newblock \bibinfo{title}{Quantum finance: Path integrals and Hamiltonians for
  options and interest rates}.
\newblock \bibinfo{publisher}{Cambridge University Press},
  \bibinfo{address}{Cambridge UK}.
\bibitem[{Bailey and Eichler(2006)}]{BaileyEichler}
\bibinfo{author}{Bailey, J.A.}, \bibinfo{author}{Eichler, E.E.},
  \bibinfo{year}{2006}.
\newblock \bibinfo{title}{Primate segmental duplications: crucibles of
  evolution, diversity, and disease}.
\newblock \bibinfo{journal}{Nature Reviews Genetics} \bibinfo{volume}{7},
  \bibinfo{pages}{552--564}.
\bibitem[{Barab\'asi and Oltvai(2004)}]{BarabasiOltvai}
\bibinfo{author}{Barab\'asi, A.}, \bibinfo{author}{Oltvai, Z.N.},
  \bibinfo{year}{2004}.
\newblock \bibinfo{title}{Network biology: Understanding the cell's functional
  organization}.
\newblock \bibinfo{journal}{Nature Reviews Genetics} \bibinfo{volume}{5},
  \bibinfo{pages}{101--113}.
\bibitem[{Baum and Donoghue(2001)}]{BaumDonoghue}
\bibinfo{author}{Baum, D.A.}, \bibinfo{author}{Donoghue, M.J.},
  \bibinfo{year}{2001}.
\newblock \bibinfo{title}{A likelihood framework for the phylogenetic analysis
  of adaptation}, in: \bibinfo{editor}{Orzack, S.H.}, \bibinfo{editor}{Sober,
  E.} (Eds.), \bibinfo{booktitle}{Adaptationism and Optimality}.
  \bibinfo{publisher}{Cambridge University Press}, \bibinfo{address}{Cambridge,
  UK}, pp. \bibinfo{pages}{24--44}.
\bibitem[{Bentley et~al.(2004)Bentley, Hahn and Shennan}]{Bentleyetal04}
\bibinfo{author}{Bentley, R.A.}, \bibinfo{author}{Hahn, M.W.},
  \bibinfo{author}{Shennan, S.J.}, \bibinfo{year}{2004}.
\newblock \bibinfo{title}{Random drift and cultural change}, in:
  \bibinfo{booktitle}{Proceedings of the Royal Society B Biological Sciences},
  pp. \bibinfo{pages}{1143--1450}.
\bibitem[{Bongard(2008)}]{Bongard}
\bibinfo{author}{Bongard, J.}, \bibinfo{year}{2008}.
\newblock \bibinfo{title}{Behavior chaining: Incremental behavior integration
  for evolutionary robotics}, in: \bibinfo{booktitle}{Artificial Life XI},
  \bibinfo{address}{Winchester, UK}.
\bibitem[{Boyd and Richerson(1985)}]{BoydRicherson85}
\bibinfo{author}{Boyd, R.}, \bibinfo{author}{Richerson, P.J.},
  \bibinfo{year}{1985}.
\newblock \bibinfo{title}{Culture and the evolutionary process}.
\newblock \bibinfo{publisher}{University of Chicago Press},
  \bibinfo{address}{Chicago, IL}.
\bibitem[{Boyd and Richerson(2005)}]{BoydRicherson05}
\bibinfo{author}{Boyd, R.}, \bibinfo{author}{Richerson, P.J.},
  \bibinfo{year}{2005}.
\newblock \bibinfo{title}{The origin and evolution of culture}.
\newblock \bibinfo{publisher}{Oxford University Press},
  \bibinfo{address}{Oxford}.
\bibitem[{Bruza et~al.(2009)Bruza, Busemeyer and
  Gabora}]{BruzaBusemeyerGabora09}
\bibinfo{author}{Bruza, P.}, \bibinfo{author}{Busemeyer, J.},
  \bibinfo{author}{Gabora, L.}, \bibinfo{year}{2009}.
\newblock \bibinfo{title}{Introduction to the special issue on quantum
  cognition}.
\newblock \bibinfo{journal}{Journal of Mathematical Psychology}
  \bibinfo{volume}{53}, \bibinfo{pages}{303--305}.
\bibitem[{Bruza and Cole(2005)}]{BruzaCole05}
\bibinfo{author}{Bruza, P.}, \bibinfo{author}{Cole, R.}, \bibinfo{year}{2005}.
\newblock \bibinfo{title}{Quantum logic of semantic space: An exploratory
  investigation of context effects in practical reasoning}, in:
  \bibinfo{editor}{Artemov, S.}, \bibinfo{editor}{Barringer, H.},
  \bibinfo{editor}{d'Avila Garcez, A.S.}, \bibinfo{editor}{Lamb, L.C.},
  \bibinfo{editor}{Woods, J.} (Eds.), \bibinfo{booktitle}{We Will Show Them:
  Essays in Honour of Dov Gabbay}. \bibinfo{publisher}{College Publications}.
  volume~\bibinfo{volume}{1}, pp. \bibinfo{pages}{339--361}.
\bibitem[{Busemeyer et~al.(2011)Busemeyer, Pothos, Franco and
  Trueblood}]{BusemeyerPothosFrancoTrueblood11}
\bibinfo{author}{Busemeyer, J.R.}, \bibinfo{author}{Pothos, E.},
  \bibinfo{author}{Franco, R.}, \bibinfo{author}{Trueblood, J.S.},
  \bibinfo{year}{2011}.
\newblock \bibinfo{title}{A quantum theoretical explanation for probability
  judgment `errors'}.
\newblock \bibinfo{journal}{Psychological Review} \bibinfo{volume}{118},
  \bibinfo{pages}{193--218}.
\bibitem[{Busemeyer et~al.(2006)Busemeyer, Wang and
  Townsend}]{BusemeyerWangTownsend06A}
\bibinfo{author}{Busemeyer, J.R.}, \bibinfo{author}{Wang, Z.},
  \bibinfo{author}{Townsend, J.T.}, \bibinfo{year}{2006}.
\newblock \bibinfo{title}{Quantum dynamics of human decision-making}.
\newblock \bibinfo{journal}{Journal of Mathematical Psychology}
  \bibinfo{volume}{50}, \bibinfo{pages}{220--241}.
\bibitem[{Carroll et~al.(2001)Carroll, Grenier and
  Weatherbee}]{CarrollGrenierWeatherbee}
\bibinfo{author}{Carroll, S.B.}, \bibinfo{author}{Grenier, J.K.},
  \bibinfo{author}{Weatherbee, S.D.}, \bibinfo{year}{2001}.
\newblock \bibinfo{title}{From DNA to Diversity: Molecular Genetics and the
  Evolution of Animal Design}.
\newblock \bibinfo{publisher}{Blackwell Science}, \bibinfo{address}{Malden,
  MA}.
\bibitem[{Cavalli and Feldman(1981)}]{Cavalli-SforzaFeldman81}
\bibinfo{author}{Cavalli, L.L.}, \bibinfo{author}{Feldman, M.W.},
  \bibinfo{year}{1981}.
\newblock \bibinfo{title}{Cultural transmission and evolution: A quantitative
  approach}.
\newblock \bibinfo{publisher}{Princeton University Press},
  \bibinfo{address}{Princeton, NJ}.
\bibitem[{Chothia et~al.(2003)Chothia, Gough, Vogel and
  Teichmann}]{ChothiaEtAl}
\bibinfo{author}{Chothia, C.}, \bibinfo{author}{Gough, J.},
  \bibinfo{author}{Vogel, C.}, \bibinfo{author}{Teichmann, S.A.},
  \bibinfo{year}{2003}.
\newblock \bibinfo{title}{Evolution of the protein repertoire}.
\newblock \bibinfo{journal}{Science} \bibinfo{volume}{300},
  \bibinfo{pages}{1701--1703}.
\bibitem[{Daniels et~al.(2004)Daniels, Orgeig, Sullivan, Ling, Bennett,
  Sch\"urch, Val and Brauner}]{DanielsEtAl04}
\bibinfo{author}{Daniels, C.B.}, \bibinfo{author}{Orgeig, S.},
  \bibinfo{author}{Sullivan, L.C.}, \bibinfo{author}{Ling, N.},
  \bibinfo{author}{Bennett, M.B.}, \bibinfo{author}{Sch\"urch, S.},
  \bibinfo{author}{Val, A.L.}, \bibinfo{author}{Brauner, C.J.},
  \bibinfo{year}{2004}.
\newblock \bibinfo{title}{The origin and evolution of the surfactant system in
  fish: Insights into the evolution of lungs and swim bladders}.
\newblock \bibinfo{journal}{Physiological and Biochemical Zoology}
  \bibinfo{volume}{77}, \bibinfo{pages}{732--749}.
\bibitem[{Darwin(1859)}]{Darwin}
\bibinfo{author}{Darwin, C.}, \bibinfo{year}{1859}.
\newblock \bibinfo{title}{The Origin of Species}.
\bibitem[{Dorigo and Colombetti(1998)}]{DorigoColombetti}
\bibinfo{author}{Dorigo, M.}, \bibinfo{author}{Colombetti, M.},
  \bibinfo{year}{1998}.
\newblock \bibinfo{title}{Robot Shaping: An Experiment in Behavior
  Engineering}.
\newblock \bibinfo{publisher}{MIT Press}, \bibinfo{address}{Cambridge, MA}.
\bibitem[{Durham(1991)}]{Durham91}
\bibinfo{author}{Durham, W.}, \bibinfo{year}{1991}.
\newblock \bibinfo{title}{Coevolution: Genes, culture and human diversity}.
\newblock \bibinfo{publisher}{Stanford University Press},
  \bibinfo{address}{Stanford}.
\bibitem[{Eichler(2001)}]{Eichler}
\bibinfo{author}{Eichler, E.E.}, \bibinfo{year}{2001}.
\newblock \bibinfo{title}{Recent duplication, domain accretion and the dynamic
  mutation of the human genome}.
\newblock \bibinfo{journal}{Trends in Genetics} \bibinfo{volume}{17}.
\bibitem[{Fentress(2005)}]{Fentress}
\bibinfo{author}{Fentress, S.W.}, \bibinfo{year}{2005}.
\newblock \bibinfo{title}{Exaptation as a means of evolving complex solutions}.
\newblock Master's thesis. University of Edinburgh.
\bibitem[{Finn et~al.(2010)Finn, mistry, Tate, Coggill, Heger, Pollington,
  Gavin, Gunesekaran, Ceric, Forslund, Holm, Sonnhammer, Eddy and
  Bateman}]{FinnEtAl}
\bibinfo{author}{Finn, R.D.}, \bibinfo{author}{mistry, J.},
  \bibinfo{author}{Tate, J.}, \bibinfo{author}{Coggill, P.},
  \bibinfo{author}{Heger, A.}, \bibinfo{author}{Pollington, J.E.},
  \bibinfo{author}{Gavin, O.L.}, \bibinfo{author}{Gunesekaran, P.},
  \bibinfo{author}{Ceric, G.}, \bibinfo{author}{Forslund, K.},
  \bibinfo{author}{Holm, L.}, \bibinfo{author}{Sonnhammer, E.L.},
  \bibinfo{author}{Eddy, S.R.}, \bibinfo{author}{Bateman, A.},
  \bibinfo{year}{2010}.
\newblock \bibinfo{title}{The pfam protein families database}.
\newblock \bibinfo{journal}{Nucleic Acids Research} \bibinfo{volume}{38},
  \bibinfo{pages}{D211--D222}.
\bibitem[{Gabora(1995)}]{Gabora95}
\bibinfo{author}{Gabora, L.}, \bibinfo{year}{1995}.
\newblock \bibinfo{title}{Meme and variations: A computational model of
  cultural evolution}, in: \bibinfo{booktitle}{Lectures in Complex Systems}.
  \bibinfo{publisher}{Addison-Wesley}, \bibinfo{address}{Reading, MA}.
\bibitem[{Gabora(1996)}]{Gabora96}
\bibinfo{author}{Gabora, L.}, \bibinfo{year}{1996}.
\newblock \bibinfo{title}{A day in the life of a meme}.
\newblock \bibinfo{journal}{Philosophica} \bibinfo{volume}{57},
  \bibinfo{pages}{901--938}.
\bibitem[{Gabora(2001)}]{Gabora01}
\bibinfo{author}{Gabora, L.}, \bibinfo{year}{2001}.
\newblock \bibinfo{title}{Cognitive mechanisms unerlying the origin and
  evolution of culture}.
\newblock Ph.D. thesis. Free University of Brussels.
\bibitem[{Gabora(2003)}]{Gabora03}
\bibinfo{author}{Gabora, L.}, \bibinfo{year}{2003}.
\newblock \bibinfo{title}{Contextual focus: A cognitive explanation for the
  cultural transition of the middle/upper paleolithic}, in:
  \bibinfo{booktitle}{Proceedings of the 25th Annual Meeting of the Cognitive
  Science Society}, \bibinfo{publisher}{Lawrence Erlbaum Associates},
  \bibinfo{address}{Boston MA}. pp. \bibinfo{pages}{432--437}.
\bibitem[{Gabora(2008)}]{Gabora08}
\bibinfo{author}{Gabora, L.}, \bibinfo{year}{2008}.
\newblock \bibinfo{title}{The cultural evolution of socially situated
  cognition}.
\newblock \bibinfo{journal}{Cognitive Systems Research} \bibinfo{volume}{9},
  \bibinfo{pages}{104--113}.
\bibitem[{Gabora(2010)}]{Gabora10}
\bibinfo{author}{Gabora, L.}, \bibinfo{year}{2010}.
\newblock \bibinfo{title}{Revenge of the 'neurds': Characterizing creative
  thought in terms of the structure and dynamics of human memory}.
\newblock \bibinfo{journal}{Creativity Research Journal} \bibinfo{volume}{22},
  \bibinfo{pages}{1--13}.
\bibitem[{Gabora(2012)}]{Gabora12}
\bibinfo{author}{Gabora, L.}, \bibinfo{year}{2012}.
\newblock \bibinfo{title}{An evolutionary framework for culture and creativity:
  Selectionism versus communal exchange}.
\newblock \bibinfo{journal}{Physics of Life Reviews} .
\bibitem[{Gabora and Aerts(2002)}]{GaboraAerts02}
\bibinfo{author}{Gabora, L.}, \bibinfo{author}{Aerts, D.},
  \bibinfo{year}{2002}.
\newblock \bibinfo{title}{Contextualizing concepts using a mathematical
  generalization of the quantum formalism}.
\newblock \bibinfo{journal}{Journal of Experimental and Theoretical Artificial
  Intelligence} \bibinfo{volume}{14}, \bibinfo{pages}{327--358}.
\bibitem[{Gabora and Aerts(2005)}]{GaboraAerts05b}
\bibinfo{author}{Gabora, L.}, \bibinfo{author}{Aerts, D.},
  \bibinfo{year}{2005}.
\newblock \bibinfo{title}{Evolution as context-driven actualization of
  potential: Toward an interdisciplinary theory of change of state}.
\newblock \bibinfo{journal}{Interdisciplinary Science Reviews}
  \bibinfo{volume}{30}, \bibinfo{pages}{69--88}.
\bibitem[{Gabora and Aerts(2008)}]{GaboraAerts08}
\bibinfo{author}{Gabora, L.}, \bibinfo{author}{Aerts, D.},
  \bibinfo{year}{2008}.
\newblock \bibinfo{title}{A cross-disciplinary framework for the description of
  contextually mediated change}, in: \bibinfo{editor}{Licata, I.},
  \bibinfo{editor}{Sakaji, A.} (Eds.), \bibinfo{booktitle}{Physics of Emergence
  and Organization}. \bibinfo{publisher}{World Scientific},
  \bibinfo{address}{Singapore}, pp. \bibinfo{pages}{109--134}.
\bibitem[{Gabora and Aerts(2009)}]{GaboraAerts09}
\bibinfo{author}{Gabora, L.}, \bibinfo{author}{Aerts, D.},
  \bibinfo{year}{2009}.
\newblock \bibinfo{title}{A model of the emergence and evolution of integrated
  worldviews}.
\newblock \bibinfo{journal}{Journal of Mathematical Psychology}
  \bibinfo{volume}{53}, \bibinfo{pages}{434--451}.
\bibitem[{Gabora and Ranjan(2013)}]{GaboraRanjan13}
\bibinfo{author}{Gabora, L.}, \bibinfo{author}{Ranjan, A.},
  \bibinfo{year}{2013}.
\newblock \bibinfo{title}{How insight emerges in distributed,
  content-addressable memory}, in: \bibinfo{editor}{Bristol, A.},
  \bibinfo{editor}{Vartanian, O.}, \bibinfo{editor}{Kaufman, J.} (Eds.),
  \bibinfo{booktitle}{The Neuroscience of Creativity}. \bibinfo{publisher}{MIT
  Press}, \bibinfo{address}{New York}.
\bibitem[{Gagn\'e and Spalding(2009)}]{GagneSpalding}
\bibinfo{author}{Gagn\'e, C.L.}, \bibinfo{author}{Spalding, T.L.},
  \bibinfo{year}{2009}.
\newblock \bibinfo{title}{Constituent integration during the processing of
  compound words: Does it involve the use of relational structures?}
\newblock \bibinfo{journal}{Journal of Memory and Language}
  \bibinfo{volume}{60}, \bibinfo{pages}{20--35}.
\bibitem[{Gentner(1998)}]{Gentner98}
\bibinfo{author}{Gentner, D.}, \bibinfo{year}{1998}.
\newblock \bibinfo{title}{Analogy}, in: \bibinfo{editor}{Bechtel, W.},
  \bibinfo{editor}{Graham, G.} (Eds.), \bibinfo{booktitle}{A companion to
  cognitive science}. \bibinfo{publisher}{Blackwell},
  \bibinfo{address}{Oxford}, pp. \bibinfo{pages}{107--113}.
\bibitem[{Gerhart and Kirschner(2007)}]{GerhartKirschner07}
\bibinfo{author}{Gerhart, J.}, \bibinfo{author}{Kirschner, M.},
  \bibinfo{year}{2007}.
\newblock \bibinfo{title}{The theory of facilitated variation}.
\newblock \bibinfo{journal}{Proceedings of the National Academy of Sciences}
  \bibinfo{volume}{104}, \bibinfo{pages}{8582--8589}.
\bibitem[{Gintis(2009)}]{Gintis}
\bibinfo{author}{Gintis, H.}, \bibinfo{year}{2009}.
\newblock \bibinfo{title}{Game Theory Evolving: A Problem-Centered Introduction
  to Modeling Strategic Interaction}.
\newblock \bibinfo{publisher}{Princeton University Press}.
\bibitem[{Gould and Vrba(1982)}]{GouldVrba}
\bibinfo{author}{Gould, S.J.}, \bibinfo{author}{Vrba, E.S.},
  \bibinfo{year}{1982}.
\newblock \bibinfo{title}{Exaptation -- a missing term in the science of form}.
\newblock \bibinfo{journal}{Paleobiology} \bibinfo{volume}{8},
  \bibinfo{pages}{4--15}.
\bibitem[{Graham(2008)}]{Graham}
\bibinfo{author}{Graham, L.}, \bibinfo{year}{2008}.
\newblock \bibinfo{title}{Exaptation and Functional Shift in Evolutionary
  Computing}.
\newblock Ph.D. thesis. Carleton University.
\bibitem[{Grover(1997)}]{Grover97}
\bibinfo{author}{Grover, L.K.}, \bibinfo{year}{1997}.
\newblock \bibinfo{title}{Quantum mechanics helps in searching for a needle in
  a haystack}.
\newblock \bibinfo{journal}{Physical Review Letters} \bibinfo{volume}{79},
  \bibinfo{pages}{325--328}.
\bibitem[{Hampton(1988)}]{Hampton88}
\bibinfo{author}{Hampton, J.A.}, \bibinfo{year}{1988}.
\newblock \bibinfo{title}{Overextension of conjunctive concepts: Evidence for a
  unitary model for concept typicality and class inclusion}.
\newblock \bibinfo{journal}{Journal of Experimental Psychology: Learning,
  Memory, and Cognition} \bibinfo{volume}{14}, \bibinfo{pages}{12--32}.
\bibitem[{Jablonka and Lamb(2005)}]{JablonkaLamb05}
\bibinfo{author}{Jablonka, E.}, \bibinfo{author}{Lamb, M.},
  \bibinfo{year}{2005}.
\newblock \bibinfo{title}{Evolution in four dimensions}.
\newblock \bibinfo{publisher}{MIT Press}.
\bibitem[{Jacob(1977)}]{Jacob}
\bibinfo{author}{Jacob, F.}, \bibinfo{year}{1977}.
\newblock \bibinfo{title}{Evolution and tinkering}.
\newblock \bibinfo{journal}{Science} \bibinfo{volume}{196},
  \bibinfo{pages}{1161--1166}.
\bibitem[{Kauffman(2008)}]{Kauffman08}
\bibinfo{author}{Kauffman, S.}, \bibinfo{year}{2008}.
\newblock \bibinfo{title}{Reinventing the Sacred: A New View of Science, Reason
  and Religion}.
\newblock \bibinfo{publisher}{Basic Books}, \bibinfo{address}{New York}.
\bibitem[{Kitto and Kortschak(2012)}]{KittoKortschak}
\bibinfo{author}{Kitto, K.}, \bibinfo{author}{Kortschak, R.D.},
  \bibinfo{year}{2012}.
\newblock \bibinfo{title}{Contextual models and the non-newtonian paradigm}.
\newblock \bibinfo{howpublished}{This volume}.
\bibitem[{Kitto et~al.(2011)Kitto, Ramm, Sitbon and Bruza}]{KittoEtAl11}
\bibinfo{author}{Kitto, K.}, \bibinfo{author}{Ramm, B.},
  \bibinfo{author}{Sitbon, L.}, \bibinfo{author}{Bruza, P.D.},
  \bibinfo{year}{2011}.
\newblock \bibinfo{title}{Quantum theory beyond the physical: information in
  context}.
\newblock \bibinfo{journal}{Axiomathes} \bibinfo{volume}{21},
  \bibinfo{pages}{331--345}.
\bibitem[{Lenski et~al.(2003)Lenski, Ofria, Pennock,  and Adami}]{Lenski}
\bibinfo{author}{Lenski, R.E.}, \bibinfo{author}{Ofria, C.},
  \bibinfo{author}{Pennock, R.T.}, , \bibinfo{author}{Adami, C.},
  \bibinfo{year}{2003}.
\newblock \bibinfo{title}{The evolutionary origin of complex features}.
\newblock \bibinfo{journal}{Nature} \bibinfo{volume}{423}.
\bibitem[{Long et~al.(2003)Long, Betr/`an, Thornton and Wang}]{LongEtAl03}
\bibinfo{author}{Long, M.}, \bibinfo{author}{Betr/`an, E.},
  \bibinfo{author}{Thornton, K.}, \bibinfo{author}{Wang, W.},
  \bibinfo{year}{2003}.
\newblock \bibinfo{title}{The origin of new genes: Glimpses from the young and
  old}.
\newblock \bibinfo{journal}{Nature Reviews Genetics} \bibinfo{volume}{4},
  \bibinfo{pages}{865--875}.
\bibitem[{Longo et~al.(2012)Longo, Mont\'evil and Kauffman}]{Longoetal12}
\bibinfo{author}{Longo, G.}, \bibinfo{author}{Mont\'evil, M.},
  \bibinfo{author}{Kauffman, S.}, \bibinfo{year}{2012}.
\newblock \bibinfo{title}{No entailing laws, but enablement in the evolution of
  the biosphere}, in: \bibinfo{booktitle}{Proceedings of the Fourteenth
  International Conference on Genetic and Evolutionary Computation Conference
  Companion}, pp. \bibinfo{pages}{1379--1392}.
\bibitem[{MacLeod(2001)}]{MacLeod}
\bibinfo{author}{MacLeod, N.}, \bibinfo{year}{2001}.
\newblock \bibinfo{title}{The role of phylogeny in quantitative paleobiological
  data}.
\newblock \bibinfo{journal}{Paleobiology} \bibinfo{volume}{27},
  \bibinfo{pages}{226--240}.
\bibitem[{Martins(2000)}]{Martins}
\bibinfo{author}{Martins, E.P.}, \bibinfo{year}{2000}.
\newblock \bibinfo{title}{Adaptation and the comparative method}.
\newblock \bibinfo{journal}{Trends in Ecology and Evolution}
  \bibinfo{volume}{15}, \bibinfo{pages}{296--299}.
\bibitem[{Mesoudi et~al.(2006)Mesoudi, Whiten and
  Laland}]{MesoudiWhitenLaland06}
\bibinfo{author}{Mesoudi, A.}, \bibinfo{author}{Whiten, A.},
  \bibinfo{author}{Laland, K.}, \bibinfo{year}{2006}.
\newblock \bibinfo{title}{Towards a unified science of cultural evolution}.
\newblock \bibinfo{journal}{Behavioral and Brain Sciences}
  \bibinfo{volume}{29}, \bibinfo{pages}{329--383}.
\bibitem[{Mithen(1996)}]{Mithen96}
\bibinfo{author}{Mithen, S.}, \bibinfo{year}{1996}.
\newblock \bibinfo{title}{The prehistory of mind: A search for the origins of
  art, science, and religion}.
\newblock \bibinfo{publisher}{Thames and Hudson}, \bibinfo{address}{London UK}.
\bibitem[{Mouret and Doncieux(2009a)}]{Mouret}
\bibinfo{author}{Mouret, J.B.}, \bibinfo{author}{Doncieux, S.},
  \bibinfo{year}{2009}a.
\newblock \bibinfo{title}{Evolving modular neural-networks through exaptation},
  in: \bibinfo{booktitle}{Eleventh IEEE Congress on Evolutionary Computation},
  \bibinfo{address}{Trondheim, Norway}.
\bibitem[{Mouret and Doncieux(2009b)}]{Mouret2}
\bibinfo{author}{Mouret, J.B.}, \bibinfo{author}{Doncieux, S.},
  \bibinfo{year}{2009}b.
\newblock \bibinfo{title}{Overcoming the bootstrap problem in evolutionary
  robotics using behavioral diversity}, in: \bibinfo{booktitle}{Eleventh
  Congress on Evolutionary Computation}, \bibinfo{address}{Trondheim, Norway}.
\bibitem[{Nielsen and Chuang(2010)}]{NielsenChuang00}
\bibinfo{author}{Nielsen, M.A.}, \bibinfo{author}{Chuang, I.L.},
  \bibinfo{year}{2010}.
\newblock \bibinfo{title}{Quantum computation and quantum information}.
\newblock \bibinfo{publisher}{Cambridge University Press},
  \bibinfo{address}{Cambridge UK}.
\bibitem[{Nowak(2006)}]{Nowak}
\bibinfo{author}{Nowak, M.}, \bibinfo{year}{2006}.
\newblock \bibinfo{title}{Evolutionary Dynamics: Exploring the Equations of
  Life}.
\newblock \bibinfo{publisher}{Belknap Press}.
\bibitem[{Oppacher and Wineberg(1999)}]{Oppacher}
\bibinfo{author}{Oppacher, F.}, \bibinfo{author}{Wineberg, M.},
  \bibinfo{year}{1999}.
\newblock \bibinfo{title}{The shifting balance genetic algorithm: Improving the
  ga in a dynamic environment}, in: \bibinfo{editor}{Banzhaf, W.},
  \bibinfo{editor}{Daida, J.}, \bibinfo{editor}{Eiben, A.E.},
  \bibinfo{editor}{Garzon, M.H.}, \bibinfo{editor}{Honavar, V.},
  \bibinfo{editor}{Jakiela, M.}, \bibinfo{editor}{Smith, R.E.} (Eds.),
  \bibinfo{booktitle}{Proceedings of the Genetic and Evolutionary Computation
  Conference}, pp. \bibinfo{pages}{603--611}.
\bibitem[{Orengo and Thornton(2005)}]{OrengoThornton}
\bibinfo{author}{Orengo, C.A.}, \bibinfo{author}{Thornton, J.M.},
  \bibinfo{year}{2005}.
\newblock \bibinfo{title}{Protein families and their evolution -- a structural
  perspective}.
\newblock \bibinfo{journal}{Annual Review of Biochemistry}
  \bibinfo{volume}{74}, \bibinfo{pages}{867--900}.
\bibitem[{Orzack and Sober(2001)}]{OrzackSober}
\bibinfo{author}{Orzack, S.H.}, \bibinfo{author}{Sober, E.},
  \bibinfo{year}{2001}.
\newblock \bibinfo{title}{Adaptationism and Optimality}.
  \bibinfo{publisher}{Cambridge University Press}. chapter
  \bibinfo{chapter}{Introduction}.
\bibitem[{Osherson and Smith(1981a)}]{OshersonSmith81}
\bibinfo{author}{Osherson, D.}, \bibinfo{author}{Smith, E.},
  \bibinfo{year}{1981}a.
\newblock \bibinfo{title}{On the adequacy of prototype theory as a theory of
  concepts}.
\newblock \bibinfo{journal}{Cognition} \bibinfo{volume}{9},
  \bibinfo{pages}{35--58}.
\bibitem[{Osherson and Smith(1981b)}]{Osherson81}
\bibinfo{author}{Osherson, D.}, \bibinfo{author}{Smith, E.},
  \bibinfo{year}{1981}b.
\newblock \bibinfo{title}{On the adequacy of prototype theory as a theory of
  concepts}.
\newblock \bibinfo{journal}{Cognition} \bibinfo{volume}{9},
  \bibinfo{pages}{35--58}.
\bibitem[{Pan and Yang(2010)}]{PanYang}
\bibinfo{author}{Pan, S.}, \bibinfo{author}{Yang, Q.}, \bibinfo{year}{2010}.
\newblock \bibinfo{title}{A survey on transfer learning}.
\newblock \bibinfo{journal}{IEEE Transactions on Knowledge and Data
  Engineering} \bibinfo{volume}{22}, \bibinfo{pages}{1345--1359}.
\bibitem[{Parter et~al.(2008)Parter, Kashtan and Alon}]{ParterKashtan}
\bibinfo{author}{Parter, M.}, \bibinfo{author}{Kashtan, N.},
  \bibinfo{author}{Alon, U.}, \bibinfo{year}{2008}.
\newblock \bibinfo{title}{Facilitated variation: How evolution learns from past
  environments to generalize to new environments}.
\newblock \bibinfo{journal}{PLoS Computational Biology} \bibinfo{volume}{4}.
\bibitem[{Patthy(2003)}]{Patthy}
\bibinfo{author}{Patthy, L.}, \bibinfo{year}{2003}.
\newblock \bibinfo{title}{Modular assembly of genes and the evolution of new
  functions}.
\newblock \bibinfo{journal}{Genetica} \bibinfo{volume}{118}.
\bibitem[{Perry et~al.(2001)Perry, Wilson, Straus, harris and
  Remmers}]{PerryEtAl01}
\bibinfo{author}{Perry, S.F.}, \bibinfo{author}{Wilson, R.J.A.},
  \bibinfo{author}{Straus, C.}, \bibinfo{author}{harris, M.B.},
  \bibinfo{author}{Remmers, J.E.}, \bibinfo{year}{2001}.
\newblock \bibinfo{title}{Which came first, the lung or the breath?}
\newblock \bibinfo{journal}{Comparative Biochemistry and Physiology Part A}
  \bibinfo{volume}{129}, \bibinfo{pages}{37--47}.
\bibitem[{Pothos and Busemeyer(2012)}]{PothosBusemeyer12}
\bibinfo{author}{Pothos, E.}, \bibinfo{author}{Busemeyer, J.R.},
  \bibinfo{year}{2012}.
\newblock \bibinfo{title}{Can quantum probability provide a new direction for
  cognitive modeling?}
\newblock \bibinfo{journal}{Behavioral and Brain Sciences} .
\bibitem[{Pothos and Busemeyer(In press)}]{BusemeyerWangTownsend06B}
\bibinfo{author}{Pothos, E.}, \bibinfo{author}{Busemeyer, J.R.},
  \bibinfo{year}{In press}.
\newblock \bibinfo{title}{Can quantum probability provide a new direction for
  cognitive modeling?}
\newblock \bibinfo{journal}{Behavioral and Brain Sciences} .
\bibitem[{Prum and Brush(2002)}]{PrumBrush}
\bibinfo{author}{Prum, R.O.}, \bibinfo{author}{Brush, A.H.},
  \bibinfo{year}{2002}.
\newblock \bibinfo{title}{The evolutionary origin and diversification of
  feathers}.
\newblock \bibinfo{journal}{The Quarterly Review of Biology}
  \bibinfo{volume}{77}, \bibinfo{pages}{261--295}.
\bibitem[{Simeonov et~al.(2011)Simeonov, Ehresmann, Smith, Ramirez and
  Repa}]{Simeonovetal11}
\bibinfo{author}{Simeonov, P.L.}, \bibinfo{author}{Ehresmann, A.C.},
  \bibinfo{author}{Smith, L.S.}, \bibinfo{author}{Ramirez, J.G.},
  \bibinfo{author}{Repa, V.}, \bibinfo{year}{2011}.
\newblock \bibinfo{title}{A new biology: A modern perspective on the challenge
  of closing the cap between the islands of knowledge}.
\newblock \bibinfo{journal}{Lecture Notes in Computer Science}
  \bibinfo{volume}{6569}, \bibinfo{pages}{188--195}.
\bibitem[{Skolicki(2007)}]{Skolicki}
\bibinfo{author}{Skolicki, Z.M.}, \bibinfo{year}{2007}.
\newblock \bibinfo{title}{An Analysis of Island Models in Evolutionary
  Computation}.
\newblock Ph.D. thesis. George Mason University.
\bibitem[{Smith(2010)}]{Smith}
\bibinfo{author}{Smith, S.D.}, \bibinfo{year}{2010}.
\newblock \bibinfo{title}{Using phylogenetics to detect pollinator-mediated
  floral evolution}.
\newblock \bibinfo{journal}{New Phytologist} \bibinfo{volume}{188},
  \bibinfo{pages}{345--363}.
\bibitem[{Taylor and Raes(2004)}]{TaylorRaes}
\bibinfo{author}{Taylor, J.S.}, \bibinfo{author}{Raes, J.},
  \bibinfo{year}{2004}.
\newblock \bibinfo{title}{Duplication and divergence: The evolution of new
  genes and old ideas}.
\newblock \bibinfo{journal}{Annual Review of Genetics} \bibinfo{volume}{38},
  \bibinfo{pages}{615--43}.
\bibitem[{Taylor and Stone(2009)}]{TaylorStone}
\bibinfo{author}{Taylor, M.E.}, \bibinfo{author}{Stone, P.},
  \bibinfo{year}{2009}.
\newblock \bibinfo{title}{Transfer learning for reinforcement learning domains:
  A survey}.
\newblock \bibinfo{journal}{Journal of Machine Learning Research}
  \bibinfo{volume}{10}, \bibinfo{pages}{1633--1685}.
\bibitem[{Torrey and Shavlik(2009)}]{TorreyShavlik}
\bibinfo{author}{Torrey, L.}, \bibinfo{author}{Shavlik, J.},
  \bibinfo{year}{2009}.
\newblock \bibinfo{title}{Transfer learning}, in: \bibinfo{editor}{Soria, E.},
  \bibinfo{editor}{Martin, J.}, \bibinfo{editor}{Magdalena, R.},
  \bibinfo{editor}{Martinez, M.}, \bibinfo{editor}{Serrano, A.} (Eds.),
  \bibinfo{booktitle}{Handbook of Research on Machine Learning Applications},
  \bibinfo{publisher}{IGI Global}.

\end{thebibliography}

\end{document}